\documentclass[aps,prl,twocolumn,groupedaddress,showpacs]{revtex4}
\usepackage{graphicx,epsf,color,amsmath}
\newif\ifdraft
\drafttrue
\newif\ifpreprint
\preprinttrue

\def\fig#1{Fig.~{\ref{#1}}}

\def\spa#1.#2{\left\langle#1\,#2\right\rangle}
\def\spb#1.#2{\left[#1\,#2\right]}
\def\tree{{\rm tree}}

\def\eps{\epsilon}

\def\eqn#1{Eq.~(\ref{#1})}

\def\NeqFour{{{\cal N}=4}}

\def\NeqEight{{{\cal N}=8}}
\def\N{{\cal N}}
\def\fourloop{{\rm 4\hbox{-}loop}}

\def\be{\begin{equation}}
\def\ee{\end{equation}}
\def\bea{\begin{eqnarray}}
\def\eea{\end{eqnarray}}
\def\ba{\begin{eqnarray}}
\def\ea{\end{eqnarray}}

\def\tree{{\rm tree}}

\newbox\charbox
\newbox\slabox
\def\s#1{{      % Feynman slash
        \setbox\charbox=\hbox{$#1$}
        \setbox\slabox=\hbox{$/$}
        \dimen\charbox=\ht\slabox
        \advance\dimen\charbox by -\dp\slabox
        \advance\dimen\charbox by -\ht\charbox
        \advance\dimen\charbox by \dp\charbox
        \divide\dimen\charbox by 2
        \raise-\dimen\charbox\hbox to \wd\charbox{\hss/\hss}
        \llap{$#1$} }}

\def\tree{{\rm tree}}

\begin{document}

\ifpreprint
UCLA/13/TEP/107 \hfill $\null\hskip 4cm \null$  \hfill
\fi

\title{The Ultraviolet Properties of $\NeqFour$ Supergravity at Four Loops}

% For arxiv
%\title{Perturbative Quantum Gravity from Gauge Theory}

\author{Zvi~Bern${}^a$, Scott~Davies${}^a$, Tristan Dennen${}^b$,
 Alexander V. Smirnov${}^c$ and  Vladimir A. Smirnov${}^d$}
\affiliation{
${}^a$Department of Physics and Astronomy, UCLA, Los Angeles, CA
90095-1547, USA \\
${}^b$Niels Bohr International Academy and Discovery Center,
The Niels Bohr Institute, Blegdamsvej 17, DK-2100 Copenhagen
 \O, Denmark\\
${}^c$Scientific Research Computing Center, Moscow State University, 
119992 Moscow, Russia\\
${}^d$Skobeltsyn Institute of Nuclear Physics of 
  Moscow State University, 119992 Moscow, Russia
}

\begin{abstract}
We demonstrate that pure $\NeqFour$ supergravity is ultraviolet
divergent at four loops.  The form of the divergence suggests that it
is due to the rigid $U(1)$ duality-symmetry anomaly of the theory.  This
is the first known example of an ultraviolet divergence in a pure
ungauged supergravity theory in four dimensions.  We use the duality
between color and kinematics to construct the integrand 
of the four-loop four-point amplitude, whose ultraviolet
divergence is then extracted by standard integration techniques.
\end{abstract}

\pacs{04.65.+e, 11.15.Bt, 11.25.Db, 12.60.Jv \hspace{1cm}}

\maketitle

Recent years have seen enormous advances in our ability to obtain
scattering amplitudes in gauge and gravity theories.  Using these
advances we can address basic questions on the ultraviolet properties
of quantum gravity that had seemed relegated to the dustbin of
undecidable questions.  Power-counting arguments suggest that all
point-like theories of gravity should be ultraviolet divergent.
However, such arguments can be misleading if there are additional
hidden symmetries or structures.  In particular, the duality between
color and kinematics~\cite{BCJ,BCJLoop} has been shown to be
responsible for improved ultraviolet behavior in the relatively simple
two-loop case of half-maximal supergravity in five
dimensions~\cite{HalfMax5D}. This example emphasizes the importance of
carrying out more general investigations of the ultraviolet properties
of supergravity theories to ascertain the full implications of new
structures.

Pure Einstein gravity has long been known to be finite at one
loop~\cite{HooftVeltman} but divergent at two
loops~\cite{GoroffSagnotti}.  It also diverges at one loop under the
addition of generic matter~\cite{HooftVeltman,Matter}.  However, the
situation with pure ungauged supergravity is less clear. Such theories
are known not to diverge prior to three
loops~\cite{SupergravityArguments}.  The consensus from studies in the
1980s was that all supergravity theories likely diverge at three loops
(see for example, Ref.~\cite{HoweStelleReview}), although with
appropriate assumptions tighter bounds are
possible~\cite{GrasaruSiegel}.  However, it was not possible to check
these arguments until the advent of the unitarity
method~\cite{UnitarityMethod,BDDPR}.  For the most supersymmetric case
of $\NeqEight$ supergravity~\cite{N8Supergravity}, explicit
calculations have shown that the four-point amplitudes are finite at
three loops for dimensions $D<6$~\cite{GravityThree} and at four loops
for dimensions $D<11/2$~\cite{GravityFour}. These ultraviolet
cancellations were subsequently shown to be a consequence of
supersymmetry and the $E_{7(7)}$ duality symmetry of the
theory~\cite{SevenLoopGravity,VanishingVolume}. However, a $D^8 R^4$
counterterm appears to be valid under all standard symmetries, leading
to predictions of a seven-loop divergence in $\NeqEight$ supergravity
in $D = 4$.

While seven loops is at present out of reach of direct computations,
reducing the supersymmetry lowers the loop order at which nontrivial
ultraviolet cancellations can be studied. As discussed in 
ref.~\cite{VanishingVolume}, the same type of symmetry
argument used for $\NeqEight$ supergravity at seven loops also
implies the existence of an apparently valid three-loop $R^4$
counterterm in $\NeqFour$
supergravity~\cite{N4Sugra}. This suggests that pure $\NeqFour$
supergravity should diverge at three loops.  This is consistent with
speculations based on the pattern of cancellations at one loop,
suggesting that at least $\N \ge 5$ supergravity is needed to tame
ultraviolet singularities~\cite{OneLoopUnexpected}.

However, as recently demonstrated, the coefficient of the potential
three-loop four-point divergence of $\NeqFour$ supergravity actually
vanishes~\cite{N4gravThreeLoops}.  (See Ref.~\cite{VanhoveN4} for a
string-theory argument.)  Another related example is the unexpected
finiteness of the two-loop four-point amplitude of half-maximal
supergravity in five dimensions~\cite{VanhoveN4,HalfMax5D}.  By
assuming the existence of appropriate 16-supercharge superspaces, the
observed finiteness can be understood as a consequence of standard
symmetries~\cite{BHSSuperSpace}.  However, these superspaces also lead
to predictions in direct contradiction to explicit calculations when
matter multiplets are added~\cite{N4Mat}, implying that the assumption
needs to be altered.  There are also conjectures that certain
structures or hidden symmetries may play a
role~\cite{KalloshFerrara}. In any case, these examples remain
unexplained from standard symmetry considerations.  This makes it
important to investigate the next loop order. If there are no additional
cancellations at four loops beyond the ones already identified at
three loops, either in string theory or in field theory, it should
diverge~\cite{VanhoveN4,BHSSuperSpace}.

%%%%%%%%% FIGURE %%%%%%%%%%%%%%%                                              
\begin{figure}[tb]
\begin{center}
\vskip .7 cm 
\includegraphics[scale=.31]{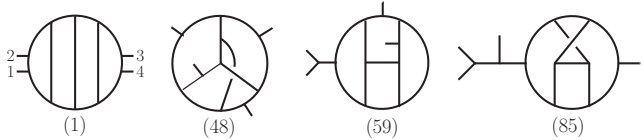}
\end{center}
\vskip -.7 cm 
\caption[a]{\small Four of the 85 diagrams with cubic vertices used 
  to organize the $\NeqFour$ super-Yang-Mills amplitudes into a form that
  respects the duality between color and kinematics.  The remaining
  diagrams are listed in Ref.~\cite{ck4l}.
\label{diagramsFigure}
}
\end{figure}
%%%%%%%%%%%%%%%%%%%%%%%%%%%%%%%%

In this Letter, we compute the four-loop four-point divergence of $\NeqFour$
supergravity following the same basic methods used in the
corresponding three-loop computation~\cite{N4gravThreeLoops} and
described in some detail in Ref.~\cite{N4Mat}.  We find that although
$\NeqFour$ supergravity does have an ultraviolet divergence, its form
suggests that it is special and tied to the $U(1)$ duality anomaly of
the theory.

Our construction of the four-loop four-point amplitude of $\NeqFour$
supergravity starts with the corresponding pure Yang-Mills Feynman
diagrams in Feynman gauge.  To obtain $\NeqFour$ supergravity, we also
need the $\NeqFour$ super-Yang-Mills diagram kinematic numerators
listed in Ref.~\cite{ck4l} that obey the duality between color and
kinematics. In this form the kinematic-numerator factors $n_i$ satisfy
algebraic relations in one-to-one correspondence with relations
satisfied by the color factors $c_i$. These factors are 
associated with 85
diagrams (plus permutations of external legs) containing only cubic
vertices, as illustrated in \fig{diagramsFigure}.  The $\NeqFour$
supergravity integrands are obtained simply by replacing the color
factors $c_i$ in the pure-Yang-Mills integrand with the corresponding
$\NeqFour$ super-Yang-Mills kinematic-numerator factors,
\begin{align}
c_i \rightarrow n_i \,. 
\end{align}
The construction of the supergravity integrand via the duality between
color and kinematics automatically satisfies the $D$-dimensional
unitarity cut constraints, given that the input gauge-theory
amplitudes have the correct cuts.  

The $\NeqFour$ super-Yang-Mills numerators~\cite{ck4l} used
in the construction are proportional to the color-ordered $\NeqFour$
super-Yang-Mills tree-level amplitudes $A^\tree_{\NeqFour}$, which can
be conveniently expressed in an on-shell superspace formalism in
four dimensions~\cite{Nair}.  As an example, diagram (1) in
\fig{diagramsFigure} has a numerator given by $n_1 = s^4 t
A^\tree_{\NeqFour}$, where $s = (k_1 + k_2)^2$ and $t = (k_2+k_3)^2$
are standard Mandelstam invariants. The remaining numerator factors
are specified in Ref.~\cite{ck4l} and are, in general, somewhat more
complicated, depending also on loop momenta. 

Using Feynman diagrams for the nonsupersymmetric pure Yang-Mills
amplitude might seem inefficient, but for the problem at hand it is a
reasonable choice. It automatically gives us local covariant
expressions with no spurious singularities that could complicate loop
integration. Moreover, only the relatively small subset of diagrams
containing color factors matching those of the nonvanishing diagrams
in the corresponding $\NeqFour$ super-Yang-Mills theory are needed,
otherwise the contribution vanishes as well in $\NeqFour$
supergravity.  Feynman diagrams also avoid subtleties associated with
the bubble-on-external-leg diagrams, such as diagram (85) of
\fig{diagramsFigure}.  After integration all such pure Yang-Mills
Feynman diagrams are smooth in the on-shell limit, canceling the
$1/k^2$ propagator as $k^2 \rightarrow 0$.
In $\NeqFour$ supergravity such contributions vanish
because the color factors in the pure Yang-Mills diagrams are replaced
by vanishing numerator factors independent of loop
momentum~\cite{ck4l}.

The logarithmic ultraviolet divergence may be extracted by series
expanding in small external momenta, or equivalently large loop
momenta~\cite{Vladimirov}.  The resulting tensor integrals are then
reduced to scalar integrals via Lorentz invariance.  We regularize the
integrals using dimensional reduction~\cite{DimRed}.  Further details
of the procedure are given in Ref.~\cite{N4Mat}.

The small-momentum expansion has the undesired effect of introducing
new unphysical infrared singularities.  To separate out all resulting
infrared divergences from the ultraviolet ones, we use a mass
regulator.  A particularly convenient choice is to introduce a uniform
mass into all Feynman propagators prior to expanding in external
momenta~\cite{ChetyrkinUniformMass}. For the case of pure 
$\NeqFour$ supergravity with no matter multiplets, with this regulator, the
subdivergences 
should all cancel amongst themselves because there are no
one-, two- or three-loop divergences.
This can be used to greatly simplify the
computation since we do not need to compute subdivergences. However,
we compute them regardless, using their cancellation as a nontrivial
consistency check.  More generally, the issue of infrared
regularization is delicate because of regulator dependence.  For
example, if the mass regulator were introduced after the expansion in
external momenta, it would ruin the cancellation of subdivergences
between different integrals, and one would need to include all
subdivergence subtractions to remove the regulator dependence.

%%%%%%%%% FIGURE %%%%%%%%%%%%%%%                                              
\begin{figure}[tb]
\begin{center}
\vskip .7 cm 
\includegraphics[scale=.4]{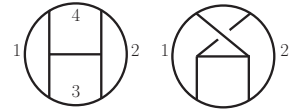}
\end{center}
\vskip -.7 cm 
\caption[a]{\small The two basic vacuum graphs.
\label{vacuumFigure}
}
\end{figure}
%%%%%%%%%%%%%%%%%%%%%%%%%%%%%%%%

At the end of this process, we obtain a large number of vacuum
integrals with the two basic diagrammatic structures shown in
\fig{vacuumFigure}.  These are of the form,
\begin{equation}
\int \prod_{j=1}^4 \frac{d^D p_j}{(2\pi)^D} 
\frac{P(m^2, p_1\cdot p_2)}{\prod_{i=1}^{9} (p_i^2 - m^2)^{a_i}} \,, 
\end{equation}
where $P$ is a numerator polynomial in the mass and the irreducible
dot product formed from the momenta flowing through propagators 1 and
2, indicated in \fig{vacuumFigure}.  (By irreducible we mean that it
cannot be expressed as a linear combination of inverse propagators and
masses.)  The 9 $p_i$ correspond to the 9 propagators in each of the
vacuum diagrams of \fig{vacuumFigure}, with the first four being
independent loop momenta.  The indices $a_i$ are integers.

The standard modern way to evaluate these vacuum integrals is to use
integration-by-parts relations \cite{ChetyrkinIBP} within dimensional
regularization.  This allows us to write down any given integral as a
linear combination of so-called master integrals which can then be evaluated. 
For four-loop Feynman vacuum integrals, this was
done in Ref.~\cite{FourLoopIntegrals}.  In our calculation, the
reduction to master integrals turns out to be complicated because high
powers of numerator loop momenta are involved. To deal with this, we
use the {\tt C++} version of the code {\tt FIRE}~\cite{FIRE},
implementing the Laporta algorithm~\cite{Laporta}.  We use the same
master-integral basis set as in Ref.~\cite{Czakon}.  (See 
Ref.~\cite{LaportaIntegrals} for a high-precision numerical
evaluation.)

Each state of pure $\NeqFour$ supergravity is a direct product of a
color-stripped state of $\NeqFour$ super-Yang-Mills theory and of pure
nonsupersymmetric Yang-Mills theory. Pure $\NeqFour$ supergravity
contains two multiplets that do not mix under linearized
supersymmetry: one contains the negative-helicity graviton and the
other the positive-helicity graviton. We
find that all amplitudes in pure $\NeqFour$ supergravity are
divergent at four loops,
\begin{equation}
\mathcal{M}^\fourloop\Bigr|_{\rm div.} = \frac{1}{(4\pi)^8} \frac{1}{\eps} 
\left(\frac{\kappa}{2}\right)^{10} \frac{1}{144} (1 -264 \zeta_3)
 \,\mathcal{T} \,,
\label{SupergravityDivergence}
\end{equation}
where $\eps = (4-D)/2$ is the dimensional-regularization parameter, and
\begin{equation}
\mathcal{T} =  s t A_{\NeqFour}^{\tree} 
\left(\mathcal{O}_1-28\mathcal{O}_2-6\mathcal{O}_3\right)\,,
\label{DivergentTensor}
\end{equation}
where
\begin{align}
\mathcal{O}_1& = \sum_{\mathcal{S}_4} \left(D_{\alpha}F_{1\mu\nu}\right)
\left(D^{\alpha}F_2^{\mu\nu}\right)F_{3\rho\sigma}F_4^{\rho\sigma}\,, \notag \\
\mathcal{O}_2& = \sum_{\mathcal{S}_4} \left(D_{\alpha}F_{1\mu\nu}\right)
\left(D^{\alpha}F_2^{\nu\sigma}\right)F_{3\sigma\rho}F_4^{\rho\mu}\,, \\
\mathcal{O}_3& = \sum_{\mathcal{S}_4} \left(D_{\alpha}F_{1\mu\nu}\right)
\left(D_{\beta}F_2^{\mu\nu}\right)F_{3\sigma}^{\hphantom{\sigma}\alpha}
  F_4^{\sigma\beta}\,. \notag
\end{align}
The sum runs over all 24 permutations of the external legs.  The
linearized field strength for each leg $j$ is given in terms of
polarization vectors for that leg,
\begin{align}
F_j^{\mu\nu}&\equiv i(k_j^{\mu}\varepsilon_j^{\nu}
-k_j^{\nu}\varepsilon_j^{\mu})\,, \notag \\
D^{\alpha}F_j^{\mu\nu}&\equiv -k_j^{\alpha}
(k_j^{\mu}\varepsilon_j^{\nu}-k_j^{\nu}\varepsilon_j^{\mu})\,.
\label{FieldStrength}
\end{align}

We have also included contributions from $\NeqFour$ matter multiplets
in the loops. As discussed in Refs.~\cite{RaduAnomaly,N4Mat}, amplitudes
with matter multiplets are straightforwardly obtained via dimensional
reduction from higher-dimensional pure half-maximal supergravity
without matter. After including the contribution of $n_{\rm V}$ matter
multiplets, with all four external states
belonging to the two graviton multiplets, the divergence is
\begin{align}
\mathcal{M}^\fourloop_{n_{\rm V} }\Bigr|_{\rm div.}  \hskip -.2 cm =
&\;\frac{1}{(4\pi)^8}\left(\frac{\kappa}{2}\right)^{10}\frac{n_{\rm V} + 2}{2304}
\Bigl[\frac{6(n_{\rm V}+2) n_{\rm V}}{\epsilon^2} \label{MatterLoops}\\
&\null 
+\frac{(n_{\rm V}+2)(3n_{\rm V}+4)
  - 96(22-n_{\rm V})\zeta_3}{\epsilon}\Bigr] \mathcal{T} . \notag
\end{align}
In this expression $n_{\rm V}$ is independent of $\eps$, a restriction
that arises from imposing this on subdivergence subtractions.  The
two- and three-loop subdivergences, and subdivergences thereof, all
cancel amongst themselves when we use a uniform mass regulator, as
happened for the $n_{\rm V} = 0$ case.  These cancellations are
analogous to similar cancellations that occur at three loops and are
surprising because there are subdivergences when matter multiplets are
included~\cite{Fischler, N4Mat}.  However, the one-loop subdivergences
do not cancel when $n_{\rm V} \not = 0$.  Instead, these enter
nontrivially to make the divergence gauge invariant and proportional
to $\mathcal{T}$.

By taking linear combinations,
\begin{align}
\mathcal{O}^{--++}&=\mathcal{O}_1-4\mathcal{O}_2\,, \hskip .8 cm 
\mathcal{O}^{-+++}=\mathcal{O}_1-4\mathcal{O}_3\,,\notag \\
\mathcal{O}^{++++}&=\mathcal{O}_2\,, 
\label{HelicityOperators}
\end{align}
each of the obtained operators are nonvanishing only for the indicated
helicity configurations and their parity conjugates and
relabelings. Here the helicity labels refer to those of the
polarization vectors used in \eqn{FieldStrength} and not the
supergravity states which are obtained by tensoring these states with
those of $\NeqFour$ super-Yang-Mills theory.  Using explicit helicity
states in $D=4$,  we have
\begin{align}
\mathcal{O}^{--++} &= 4 s^2 t \frac{\spa1.2^4}{\spa1.2\spa2.3\spa3.4\spa4.1}\,,
                      \notag\\
\mathcal{O}^{-+++} &= -12 s^2 t^2 
                 \frac{\spb2.4^2}{\spb1.2\spa2.3\spa3.4\spb4.1}\,,
                     \\
\mathcal{O}^{++++} &= 3 s t (s+t) \frac{\spb1.2\spb3.4}{\spa1.2\spa3.4}\,,
\notag
\end{align}
using spinor-helicity notation. (See Ref.~\cite{SpinorHelicity} for a
recent review.)  The divergence is thus present in all nonvanishing four-point
amplitudes of $\NeqFour$ supergravity.  Linearized supersymmetry acts
only on the $A^\tree_{\NeqFour}$ factor in \eqn{DivergentTensor}, so
each of these three configurations will {\it not} mix under this
symmetry.

The appearance of the divergences in all three independent helicity
configurations in \eqn{HelicityOperators} is surprising.  In
general, the analytic structure of amplitudes in the ${-}{-}{+}{+}$
sector is rather different from those of the other two sectors.  This
follows from generalized unitarity, where we decompose the
supergravity loops into sums of products of tree amplitudes.  In the
${-}{+}{+}{+}$ and ${+}{+}{+}{+}$ sectors, all generalized cuts vanish
in four dimensions because at least one tree amplitude will vanish.
The same does not hold in the ${-}{-}{+}{+}$ sector.  In particular,
at one loop this implies that amplitudes in the ${-}{-}{+}{+}$ sector
contain logarithms while amplitudes in the other two sectors are
pure rational functions.  The rational functions appearing in these
sectors have been directly interpreted~\cite{RaduAnomaly} as due to
the $U(1)$ duality-symmetry anomaly~\cite{MarcusAnomaly}.  We can
understand the similarity of the four-loop ultraviolet divergence in
all three sectors if we assume that it is due to the anomaly.  As
already noted in Ref.~\cite{RaduAnomaly}, unitarity implies that the
anomaly contributes to higher-loop divergences in the ${-}{-}{+}{+}$
sector as well (unless canceled from another source).  The similarity
of the divergence in all three sectors would be a consequence of
it arising from the same source.  Another helpful clue comes from the
fact that the divergence in \eqn{MatterLoops} is proportional to
$n_{\rm V} +2$.  As explained in Ref.~\cite{RaduAnomaly}, the anomaly
terms are proportional to this factor, providing further nontrivial
evidence that the four-loop divergence is due to the anomaly.

We can re-express the divergences in terms of counterterms involving
the Riemann tensor.  If we restrict the external states to four
dimensions, numerical analysis reveals that the four-external-graviton
counterterm can be reduced to a rather simple expression,
\begin{align}
C = -\frac{1}{(4\pi)^8}\left(\frac{\kappa}{2}\right)^6
\frac{1}{72 \epsilon} (1-264\zeta_3)(T_1+2 T_2)\,,
\end{align}
where
\begin{align}
\begin{array}{ll}
T_1& \equiv(D_{\alpha}R_{\mu\nu\lambda\gamma})
(D^{\alpha}R_{\rho\sigma}^{\hphantom{\rho\sigma}\lambda\gamma})R^{\nu\rho}_{\hphantom{\nu\rho}\delta\kappa}
R^{\sigma\mu\delta\kappa}\,, \\
T_2& \equiv(D_{\alpha}R_{\mu\nu\lambda\gamma})
(D^{\alpha}R_{\rho\sigma}^{\hphantom{\rho\sigma}\lambda\gamma})
R^{\mu\nu}_{\hphantom{\mu\nu}\delta\kappa}
R^{\rho\sigma\delta\kappa}\,.
\end{array}
\end{align}
Using the divergence given in \eqn{SupergravityDivergence}, one can also
obtain the explicit counterterms for any other external states of the
theory.

In any calculation of this type, it is important to have nontrivial
consistency checks on the results.  The most obvious one is the gauge
invariance of the results (\ref{SupergravityDivergence}) and
(\ref{MatterLoops}).  This requires intricate cancellations among the
terms.  We also find a required cancellation of poles in $\eps$, as
well as an expected~\cite{ChetyrkinIBP} cancellation of various
transcendental constants.  Because there are no lower-loop divergences
in pure $\NeqFour$ supergravity, only a $1/\eps$ pole can remain at four
loops.  As an illustration, consider the basis integral corresponding
to the first integral in \fig{vacuumFigure}, with all propagators
having unit indices, except for the ones labeled by 3 and 4 which have
vanishing indices (PR9 in the notation of Ref.~\cite{Czakon}). 
Up to an overall factor,
the
divergent parts of this basis integral are
\begin{align}
\null\hskip -.19 cm
{\rm PR9} &= \frac{1}{4\eps^4} + 
\frac{7}{3\eps^3} + 
\frac{1}{\eps^2}\Bigl(\frac{169}{12}-\frac{27}{2}{\rm S2} + \frac{1}{2}\zeta_2
   + \zeta_3 \Bigr) \\
& \null\hskip .4 cm  +
\frac{1}{\eps} \Bigl(\frac{143}{3} - \frac{135}{2} {\rm S2}  -  {\rm T1ep} 
+ \frac{1}{6}\zeta_2 - \frac{4}{3} \zeta_3 + \frac{3}{2} \zeta_4 \Bigr)\,,
\notag 
\end{align}
where S2 and T1ep are transcendental constants specified in
Ref.~\cite{Czakon}.  Besides finding the required cancellation of all
poles down to the $1/\eps$ level in \eqn{SupergravityDivergence}, the
transcendental constants other than $\zeta_3$ also cancel.

Another cross check on our procedure comes from computing the
coefficient of an analogous potential divergence in pure Yang-Mills
theory.  By renormalizability, the divergences are proportional to
tree-level color tensors, so all divergences containing independent
color tensors other than the tree-level ones must vanish.  Using
identical methods as for the supergravity case, we have confirmed the
ultraviolet finiteness of terms multiplying the two independent
four-loop color tensors listed in Appendix B of
Ref.~\cite{FourLoopYM}.

Instead of providing definitive answers for the ultraviolet behavior
of supergravity theories, our calculation raises additional
interesting questions.  We showed that the nonvanishing four-loop divergence
of $\NeqFour$ supergravity has a form suggesting that it is caused by
the $U(1)$ duality-symmetry anomaly. It would be important to demonstrate
this directly either via the counterterm structure or by tracking the
contributions of the anomaly to the amplitudes.  One may also wonder whether 
it is possible to remove the divergence by adding a finite term to the action so that 
an appropriate symmetry is preserved.  A key issue is to
find the higher-loop ultraviolet behavior of $\mathcal{N} \ge 5$
supergravity theories, since these should be free of duality anomalies and
therefore free of potential divergences from this source. An important
step towards this goal would be to develop improved means for
constructing representations of super-Yang-Mills amplitudes that
satisfy the duality between color and kinematics.  Another 
interesting problem is that at present there
is no complete symmetry explanation for the cancellation of the
four-point ultraviolet divergences at three loops in four-dimensional $\NeqFour$
supergravity or at two loops in five-dimensional half-maximal
supergravity.  It would be desirable to investigate this further.  If
history is any guide, further surprises await us as we probe
supergravity theories to ever deeper levels.

\vskip .1 cm

We especially thank Guillaume Bossard, Kelly Stelle and Radu Roiban for 
detailed discussions and for important comments on the manuscript. 
We also thank Simon Badger, John Joseph Carrasco, Kostja
Chetyrkin, Yu-tin Huang, Henrik Johansson, Donal O'Connell, Pierre Vanhove and 
Valery Yundin for helpful discussions. This
material is based upon work supported by the Department of Energy
under Award Numbers DE-{S}C0009937 and DE-{S}C0008279.  The work of
A.S. and V.S. was supported by the Russian Foundation for Basic
Research through grant 11-02-01196.  We thank Academic Technology
Services at UCLA for computer support.

%%%%%%%%%%%%%%%%%%%%%%%%%

\end{document}